\documentclass[aps,prd,onecolumn,groupedaddress,showpacs,nofootinbib,amssymb
]{revtex4}
\usepackage[dvips]{graphicx}
\usepackage{amssymb}
\usepackage{amsmath}
\usepackage{graphicx,,color}
\usepackage{amsfonts}
\usepackage{bm}
\usepackage{cancel}
\usepackage{comment}
\allowdisplaybreaks[4]

\begin{document}

\tolerance=5000

\newcommand\be{\begin{equation}}
\newcommand\ee{\end{equation}}
\newcommand\nn{\nonumber \\}
\newcommand\e{\mathrm{e}}

\title{Considerations on gravitational waves in higher-order local and non-local  gravity}

\author{S. Capozziello$^{1,2,3}$, M. Capriolo$^{4}$ and   S. Nojiri$^{5,6}$}

\affiliation{\it $^1$ Dipartimento
di Fisica``E. Pancini", Universit\`{a} di Napoli {}``Federico II''\\
$^2$INFN Sez. di Napoli, Compl. Univ. di Monte S. Angelo, Edificio G, Via Cinthia, I-80126, Napoli, Italy,\\
$^3$Tomsk State Pedagogical University,
634061 Tomsk, Russia,\\
$^4$Dipartimento di Matematica, Universit\`a di Salerno, Via Giovanni Paolo II, 132, Fisciano, SA I-84084, Italy.\\
$^5$ Department of Physics,
Nagoya University, Nagoya 464-8602, Japan\\
$^6$ Kobayashi-Maskawa Institute for the Origin of Particles and the Universe, Nagoya University, Nagoya 464-8602, Japan}

\date{\today}
\begin{abstract}
The detection of gravitational wave modes and polarizations could constitute an extremely important signature to discriminate among different theories of  gravity. 
According to this statement, it is possible to prove that higher-order non-local gravity  has formally the same gravitational  spectrum   of higher-order local gravity. In particular,  we consider the cases of $f \left( R, \Box R, \Box^2 R, \cdots, \Box^n R \right) 
= R + \sum_{i=1}^n \alpha_i R \Box^i R$ gravity, linear with respect to both $R$ and $\Box^i R$  and $ f \left( R, \Box R \right) 
= R + \alpha \left(\Box R\right)^2 $ gravity, quadratic with respect to $\Box R$, where it is 
demonstrated the  graviton amplitude changes  if compared with 
General Relativity.  We also obtain the  gravitational  spectrum of higher-order non-local gravity $ f \left( R, \Box^{-1} R, \Box^{-2} R, \cdots, \Box^{-n} R \right) 
= R + \sum_{i=1}^n \alpha_i R \Box^{-i} R$. In this case, we have  three state of polarization and $n+3$ oscillation modes. More in detail, it is possible to derive  two transverse tensor $(+)$ and $(\times)$ standard polarization modes of frequency $\omega_{1}$, massless and with 2-helicity; $n+1$ further scalar modes of frequency $\omega_{2},\dots,\omega_{n+2}$, massive and with 0-helicity, each of which has the same mixed polarization, partly longitudinal and partly transverse.  
\end{abstract}


\maketitle

%%%%%%%%%%%%%%%%%%%%%%%%%%%%%%%%%%%%%%%%%%%%%%%
\section{Introduction}
\label{uno}
Apart from its remarkable success to interprete today cosmological observations, the $\Lambda$-Cold Dark Matter ($\Lambda$CDM) model still lacks in according a satisfactory explanation to the issue why the energy density of the cosmological constant is so small if compared to the vacuum energy of the Standard Model (SM) of particle physics.  Furthermore, the today observed equivalence, in order of magnitude, of dark matter and dark energy escapes any  general explanation but requires the introduction of  very strict fine tunings to be addressed. 

Starting from these facts,   the cosmological constant cannot be assumed fully responsible for the whole accelerated dynamics, and one has seriously  to take into account the incapability, up today, to find  final dark matter candidates  or a self-consistent quantum theory of gravity.    According to these considerations, many scientists started questioning whether  General Relativity (GR) has  to be modified  in order to explain the accelerating expansion and the large scale structure  without  introducing  {\it "ad hoc"}  cosmological constant or new fundamental particles, (see, for example \cite{Capozziello:2011et,reportsergei,Clifton:2011jh}). The most standard modifications consist in taking into account  scalar fields either in the matter sector  or in modifying gravity. In some sense, the issue is related to adding new matter fields (dark matter, quintessence, etc.) or improving the geometry considering further degrees of freedom of gravitational field.

An important remark is necessary at this point.  Modified theories of gravity  are taken into account  to achieve a comprehensive picture of cosmic dynamics ranging from early inflation, up to large scale structure formation and  current acceleration of the universe \cite{Capozziello:2011et,reportsergei,Clifton:2011jh, Oiko}. The approach is aimed to give, in principle, a full geometric picture of cosmic history consisting, for example, in extensions of GR, like $f(R)$. The main task is explaining dynamics by further degrees of freedom of gravitational field (with respect to GR) instead of invoking dark components \cite{francaviglia}.   Furthermore, terms like $\Box R$, $\Box^2R$ and so on appear as UV corrections and  have effects also at IR scales  \cite{Capozziello:2011et}. In this perspective, further scalar fields, having a geometric  or a matter origin, could be useful to describe coherently cosmic dynamics at any scale. 

Besides these issue,   any self-consistent  theory of gravity, in order to be  renormalizable and unitary, has to face the problem of non-locality. Recently,  new approaches to overcome GR shortcomings have been related to  the breaking of  locality principle. It is worth stressing that dynamical non-locality is a property shared by all non-gravitational  fundamental interactions when  one-loop effective actions are taken into account  \cite{Barvinsky:2014lja}. 

In general, non-local extensions of GR describe  gravity by non-local effective actions. 
As demonstrated in \cite{Modesto:2011kw, Briscese:2012ys}, these kind of  theories can cure  black hole and Big Bang singularities. Furthermore, as shown   in   \cite{Deser:2007jk}, terms like  $\square^{-1}R$  could account for the late-time cosmic expansion (see also \cite{Nojiri:2007uq, Bahamonde:2017sdo}. From a fundamental physics point of view, they emerge from the seek for  renormalizable and unitary quantum theories of gravity   \cite{Biswas:2011ar} and  IR quantum corrections come out from  QFT on curved spacetime \cite{Barvinsky:2014lja}.

 With these considerations in mind, a general
class of  higher-order-scalar-tensor theories in four dimensions is
 \footnote{In this paper, we do
not consider  invariants like $R_{\mu\nu}R^{\mu\nu}$,
$R_{\mu\nu\alpha\beta}R^{\mu\nu\alpha\beta}$,
$C_{\mu\nu\alpha\beta}C^{\mu\nu\alpha\beta}$ which are also
possible.} given by the action
 \begin{eqnarray} \label{V3.1} {\cal S}=\int
d^{4}x\sqrt{-g}\left[F(R,\Box R,\Box^{2}R,..\Box^kR,\phi)
 -\frac{\epsilon}{2}
g^{\mu\nu} \phi_{; \mu} \phi_{; \nu}+ 2\kappa^2{\cal L}^{(m)}\right], \end{eqnarray} where $F$ is
a function of curvature scalar invariant $R$ and of  generic scalar
fields $\phi$. The term ${\cal L}^{(m)}$ is the minimally
coupled ordinary matter contribution; $\epsilon$ is a constant which specifies the theory. Actually its
 values can be $\epsilon =\pm 1,0$ fixing the nature and the
 dynamics of the scalar field which can be a standard scalar
 field, a phantom field or a field without dynamics \cite{Capozziello:2011et}.
In the metric formalism, the field equations are obtained by
varying (\ref{V3.1}) with respect to  $g_{\mu\nu}$.  We get
  \begin{eqnarray} 
\label{3.2cc} G^{\mu\nu}&=&\frac{1}{{\cal
G}}\left[\kappa^2 T^{\mu\nu}+\frac{1}{2}g^{\mu\nu} (F-{\cal G}R)+
(g^{\mu\lambda}g^{\nu\sigma}-g^{\mu\nu} g^{\lambda\sigma})
{\cal G}_{;\lambda\sigma}\right.\nonumber\\
& & +\frac{1}{2}\sum_{i=1}^{k}\sum_{j=1}^{i}(g^{\mu\nu}
g^{\lambda\sigma}+
  g^{\mu\lambda} g^{\nu\sigma})(\Box^{j-i})_{;\sigma}
\left(\Box^{i-j}\frac{\partial F}{\partial \Box^{i}R}\right)_{;\lambda}\nonumber\\
& &\left.-g^{\mu\nu} g^{\lambda\sigma}\left((\Box^{j-1}R)_{;\sigma}
\Box^{i-j}\frac{\partial F}{\partial \Box^{i}R}\right)_{;\lambda}\right]\,,
\end{eqnarray} where $G^{\mu\nu}$ is the  Einstein tensor and the function ${\cal G}$ is 

 \begin{eqnarray} 
\label{3.4gg}
  {\cal G}\equiv\sum_{j=0}^{n}\Box^{j}\left(\frac{\partial F}{\partial \Box^{j} R}
\right)\;. \end{eqnarray}
Eqs.(\ref{3.2cc}) are of order
$(2k+4)$. The stress-energy tensor is  
\begin{eqnarray}  \label{3.51}
T_{\mu\nu}=T^{(m)}_{\mu\nu}+\frac{\epsilon}{2}\left[\phi_{;\mu}\phi_{;\nu}-
\frac{1}{2}\phi_{;}^{\alpha}\phi_{; \alpha}\right]\;.  
\end{eqnarray}
The possible  contribution of
a potential $V(\phi)$ is contained in the definition of $F$. From now
on, we shall indicate by a capital $F$ a Lagrangian density
containing also the contribution of a potential $V(\phi)$ and by
$F(\phi)$, $f(R)$, or $f(R,\Box R)$ a function of such fields
 without scalar field potential. 
Modifications can be considered also in the framework of Teleparallel Equivalent of General Relativity (TEGR)  \cite{RepT,Li}. In such a case, dynamics is given by the torsion scalar $T$.
In  more general cases, the Lagrangian can be a function of both $T$ and $R$, \emph{i.e.}  $f(T,R)$~\cite{Myrzakulov,CapoMyr}. The connection between torsion scalar  $T$ and curvature scalar  $R$ 
is achieved by a boundary term $B$ as discussed, for example, in \cite{Wright2,noetherseb}.

The above class of models  \eqref{V3.1} is a typical example  of Extended Theories of Gravity \cite{Capozziello:2011et} where the paradigm is to recover GR as a particular case of function $F$ or in some limit $F\to R$. The same approach is valid also for TEGR extensions.

Clearly, GR extensions like $f \left( R, \Box^{-1} R, \Box^{-2} R, \cdots, \Box^{-n} R \right) $  has to be taken into account  for higher-order terms when non-locality is considered. As a consequence an approach like \eqref{V3.1} can be pursued also for non-local gravity.

In this discussion, a key role is played by gravitational waves (GWs) which can be the main tool to discriminate among competing theories both in local and in non-local description of gravity.

In both GR and TEGR,  GWs have two independent polarizations, usually denoted as  plus $(+)$ and cross $(\times)$.
In principle,  extra polarizations  should appear  in modified/extended  theories and their presence is one of the main  features capable of characterizing a given theory.
Studying perturbations  in the weak field limit is the standard approach  to find out the number and the nature of these possible further  polarizations.

Another approach is  the so-called  Newman-Penrose (NP) formalism~\cite{Eardley1, Eardley2}.
Adopting the  NP formalism in a generic metric theory,  plane GWs have six independent modes of polarization: considering the $z$-direction as the propagation direction of GWs, they are  plus $(+)$, cross $(\times)$, breathing $(b)$, longitudinal $(l)$, vector-$x$ $(x)$ and vector-y $(y)$ modes.
These modes can be described by the  independent NP quantities $ \{ \Psi_2, \Psi_3, \Psi_4, \Phi_{22}\} $, where $\Psi_3$ and $\Psi_4$ are complex and  each one describes two polarization modes.

The extra polarization modes can be used to discriminate among modified theories of gravity beyond GR (see, e.g. Ref.~\cite{ Bogdanos, Calmet}).
 As shown in \cite{Bamba}, GWs  in $f(T)$,  and in its scalar-tensor representation, are equivalent to those in GR and TEGR.
In $f(R)$ gravity, where the Lagrangian is an arbitrary function of the Ricci scalar, three  modes exist ~\cite{Alves1, Capozziello1, Rizwana} as well as in the $f(T,B)$ gravity, which is dynamically   equivalent to $f(R)$ ~\cite{CCC1}. Models like $f(R, \Theta)$ and $f(R,\Theta^\phi)$ have been  studied in Ref.~\cite{Alves2}, where $\Theta$ and $\Theta^\phi$ are the traces of the energy-momentum tensors of standard matter and of a scalar field, respectively. Useful results  occurs if one  calculates the gravitational energy-momentum tensor for  $f\left(R\right)$ and $f\left(T\right)$ models  and for the  Lagrangian $L_{\Box^{k}R}=\sqrt{-g}\left(\bar{R}+\sum_{k=0}^{p}a_{k}R\Box^{k}R\right)$. See, for example,  \cite{CCT2, CCT1}.

According to these considerations, GWs detection can play an important role in selecting viable theories of gravity. In fact, as discussed in \cite{Bajardi}, the features of gravitational radiation can fully characterize a given theory of gravity considering polarizations and scalar and tensor modes. 

As  recently reported, see e.g. \cite{GW_multi, speed1, lombriser}, 
gravitational astronomy and 
multimessenger astrophysics  can contribute in selecting viable theories of gravity. An important role is played, for example, by the GWs speed  if compared with the electromagnetic counterpart of emitting sources. Any difference   can be related to massive gravitational modes. Furthermore, Extended Theories of Gravity can  imply further polarizations besides the two standard $\times$ and $+$ of GR, thanks to the further degrees of  freedom often emerging. The possible detection (or non-detection) of this further modes could be a formidable tool to discriminate among the various competing theories of gravity \cite{Taishi1,Taishi2}.  
Clearly, also non-locality could be related to GWs detection being a feature characterizing wide classes of theories.

Specifically, higher-order non-local gravity may be probed in binary black holes phenomenology. In fact near a black hole, we are in a strong gravity regime, where standard Einstein's General Relativity  has to be   improved by effective terms coming from  quantum effects.  Features of non-locality   emerge as characteristic lengths and effective massive modes.  In particular, analyzing  polarizations and modes of GWs generated by the merging  of binary black holes, one could detect  non-locality as possible massive scalar modes with zero-helicity in addition to the two standard $+$ and $\times$ polarizations. In this perspective,  we need to consider very massive black holes capable of exciting further massive scalar modes. A preliminary study in this direction is reported in  \cite{Kostas}  where  non-local gravity effects are constrained considering  S2 star orbiting around the massive black hole at the Galactic Center, SgrA$^*$. There,  deviations from the Keplerian orbits have been taken into account but non-local effects do not play a significant role.  The reason is that the weak-field regime has to  be  considered in order to study orbits for  the loose system S2 -- SgrA$^*$. In this configuration, characteristic features, due to non-local gravity, do not emerge because standard General Relativity is dominating. This is not the case in  tight systems  like binary black holes where the gravitational field is extremely strong. In other words, the emergence of GW scalar massive modes could be a self-consistent probe for non-local gravity.

In this paper, we want to compare  GWs in local and non-local theories of gravity considering higher-order derivative terms into  effective actions. As said above, these terms  have an important role   in the  loop expansions of gravitational interaction \cite{Birrell} and could give rise to further polarizations.

The paper is organized as follows: in Sect. \ref{2},  we start with the action  $f \left( R, \Box R, \Box^2 R, \cdots, \Box^n R \right)$ and introduce  scalar fields  in view of putting in evidence the further degrees of freedom   in  the Einstein frame. By variation,  we obtain the  field equations of scalar fields with related constraints and  eliminate  ghost fields  thanks to the invariance of the action under BRS transformations. In particular the case $f \left( R, \Box R, \Box^2 R, \cdots, \Box^n R \right) 
= R + \sum_{i=1}^n \alpha_i R \Box^i R$ with $n=1,2$, linear with respect to both $R$ and $\Box^i R$ and the case $ f \left( R, \Box R \right) 
= R + \alpha \left(\Box R\right)^2 $, quadratic with respect to $\Box^i R$, are examined. The presence of ghost fields is considered adopting a perturbative analysis.

 In Sect. \ref{3},   higher order non-local models $f \left( R, \Box^{-1} R, \Box^{-2} R, \cdots, \Box^{-n} R \right) $ are taken into account. Using the same procedure of Sect. \ref{2}, the spectrum of gravitational modes is derived and  compared with the analogue  in  $f \left( R, \Box R, \Box^2 R, \cdots, \Box^n R \right)$ considering analogue scalar fields. In Sect. \ref{4}, we develop GWs considerations for local and non-local gravity. Discussion and conclusions  are drawn in Sect. \ref{Conclusions}.

\section{Higher-order local gravity }\label{2}

Starting from the general action \eqref{V3.1}, we can restrict to  higher-order models considering
\begin{equation}
\label{hfr1}
S = \frac{1}{2\kappa^2} 
\int d^4 x \sqrt{-g} f \left( R, \Box R, \Box^2 R, \cdots, \Box^n R \right) \, ,
\end{equation}
where, as above,  $R$ is the Ricci curvature scalar, the coupling is $\kappa^{2}=8\pi$ in natural units  $G=c=1$, and $n$ is a positive integer. 
We can introduce   scalar fields $\phi_0$, $\phi_1$, $\cdots$ $\phi_n$, $\rho $, $\lambda_1$, 
$\cdots$, $\lambda_n$ of geometric origin so then the action (\ref{hfr1}) can be rewritten as follows,
\begin{align}
\label{hfr2}
S =& \frac{1}{2\kappa^2} \int d^4 x \sqrt{-g} \left\{ f \left( \phi_0, \phi_1, \cdots , \phi_n \right) 
+ \e^{-\rho } \left( R - \phi_0 \right) + \lambda_1 \left( \Box \phi_0 - \phi_1 \right) 
+ \lambda_2 \left( \Box \phi_1 - \phi_2 \right) \right. \nn
& \left. + \cdots 
+ \lambda_n \left( \Box \phi_{n-1} - \phi_n \right) 
\right\} \, .
\end{align}
Here $\lambda_i$ are Lagrange multipliers constraining the  scalar fields $\phi_i$ with respect to the d'Alembert operators $\Box$.
By rescaling the metric, 
\begin{equation}
\label{hfr3}
g_{\mu\nu}\to \e^{\rho } g_{\mu\nu}\, ,
\end{equation}
the action (\ref{hfr2}) can be further rewritten as follows, 
\begin{align}
\label{hfr4}
& S = \frac{1}{2\kappa^2} \int d^4 x \sqrt{-g} \Bigl\{ R 
 - \frac{3}{2} \partial_\mu \rho \partial^\mu \rho 
 - \e^{\rho } \left( \partial_\mu \lambda_1 \partial^\mu \phi_0 
+ \partial_\mu \lambda_2 \partial^\mu \phi_1 + \cdots 
+ \partial_\mu \lambda_n \partial^\mu \phi_{n-1} \right) \nn
& \qquad \qquad \qquad \qquad 
 - V \left( \phi_0, \phi_1, \cdots , \phi_n, \rho , \lambda_1, \cdots , \lambda_n \right) 
\Bigr\} \, , \nn
& V \left( \phi_0, \phi_1, \cdots , \phi_n, \rho , \lambda_1, \cdots , \lambda_n \right) 
\equiv \e^{2\rho } \left( - f \left( \phi_0, \phi_1, \cdots , \phi_n \right) 
+ \e^{-\rho } \phi_0 + \lambda_1 \phi_1 + \lambda_2 \phi_2 + \cdots 
+ \lambda_n \phi_n \right) \, ,
\end{align}
where we have defined a self-interaction potential related to the scalar fields $\phi_i$ and the Lagrange multipliers $\lambda_i$.
The variation of the action with respect to $\rho $ gives, 
\begin{equation}
\label{hfr5}
0 = 3 \Box \rho 
 - \e^{\rho } \left( \partial_\mu \lambda_1 \partial^\mu \phi_0 
+ \partial_\mu \lambda_2 \partial^\mu \phi_1 + \cdots 
+ \partial_\mu \lambda_n \partial^\mu \phi_{n-1} \right) 
 - \e^{2\rho } \left( - 2 f \left( \phi_0, \phi_1, \cdots , \phi_n \right) 
+ \e^{-\rho } \phi_0 \right) \, ,
\end{equation}
and the variations with respect to $\lambda_i$ $\left( i = 1,2, \cdots, n \right)$ gives
\begin{equation}
\label{hfr6}
0 = \nabla_\mu \left( \e^{\rho } \partial^\mu \phi_{i-1} \right) 
 - \e^{2\rho } \phi_i \, .
\end{equation}
On the other hand, from the variations with respect to $\phi_0$, $\phi_i$ 
$\left( i=1,2,\cdots,n-1 \right)$, and $\phi_n$, we obtain, 
\begin{align}
\label{hfr7}
0=& \nabla_\mu \left( \e^{\rho } \partial^\mu \lambda_1 \right) 
+ \e^{2\rho } \frac{\partial f \left( \phi_0, \phi_1, \cdots , \phi_n \right) }{\partial \phi_0} 
 - \e^{\rho } \, , \\
\label{hfr8}
0=& \nabla_\mu \left( \e^{\rho } \partial^\mu \lambda_{i+1} \right)
+ \e^{2\rho } \left( 
\frac{\partial f \left( \phi_0, \phi_1, \cdots , \phi_n \right) }{\partial \phi_i} 
+ \lambda_i \right)\, , \\
\label{hfr9}
0=& \e^{2\rho } \left( 
\frac{\partial f \left( \phi_0, \phi_1, \cdots , \phi_n \right) }{\partial \phi_n} 
+ \lambda_n \right) \, 
\end{align}
Because the derivative term of $\phi_n$ does not appear in the action (\ref{hfr4}), 
$\phi_n$ is not dynamical but just an auxiliary field. 
Here we have assumed that Eq.~(\ref{hfr9}) can be solved by $\phi_n$, but if the action 
is linear with respect to $\phi_n$, Eq.~(\ref{hfr9}) gives a constraint and the system becomes 
 a little bit complicated. Such an example is given later. 
Anyway if we assume that Eq.~(\ref{hfr9}) can be solved with $\phi_n$, 
besides the metric, the dynamical fields are $\rho$, $\lambda_i$, and 
$\phi_{i-1}$, where $i=1,2,\cdots,n$. 
In $\lambda_i$ and $\phi_{i-1}$ $\left(i=1,2,\cdots,n\right)$, half of them are canonical 
but the remaining half are ghost. 

If the matter is coupled with gravity by the minimal coupling in the Jordan frame 
where the action is given by (\ref{hfr1}), the rescaling (\ref{hfr3}) tells that the matter 
directly couples with the metric and $\rho$ is defined in  the Einstein frame where the action is 
given by (\ref{hfr4}) but the matter does not directly couples with $\lambda_i$ and $\phi_i$. 
The scalar fields $\lambda_i$ and $\phi_i$ appear via the interaction. 
Therefore if the mass of the scalar fields $\lambda_i$ and $\phi_i$ are large enough, 
the ghosts could not appear.  

Anyway in order to avoid  ghosts, we may consider the BRS transformation \cite{BRS}, 
\begin{equation}
\label{hfr10}
\delta \phi_i =0 \, , \quad \delta \lambda_i = \epsilon c_i \, , \quad 
\delta c_i = 0 \, , \quad \delta b_i = \epsilon \phi_i \, .
\end{equation}
Here $i$ runs from $0$ to $n$, $i=0,1,\cdots,n$, $\epsilon$ is an anti-commuting parameter, 
and we have introduced a new field $\lambda_0$, ghosts $c_i$, and anti-ghosts $b_i$. 
Then we find 
\begin{align}
\label{hfr11}
& \delta \left( \left( - \e^{-\rho_0} + \Box \lambda_1 \right) b_0
+ \left( - \lambda_1 + \Box \lambda_2 \right) b_1 + \cdots 
+ \left( - \lambda_{n-1} + \Box \lambda_n \right) b_{n-1}  
 - \lambda_n b_n \right) =  \epsilon \left( \mathcal{L}_\mathrm{GF} 
+ \mathcal{L}_\mathrm{ghost} \right) \, , \nn
& \mathcal{L}_\mathrm{GF} = \left( - \e^{-\rho_0} + \Box \lambda_1 \right) \phi_0
+ \left( - \lambda_1 + \Box \lambda_2 \right) \phi_1 + \cdots 
+ \left( - \lambda_{n-1} + \Box \lambda_n \right) \phi_{n-1}  
 - \lambda_n \phi_n \, , \nn
& \mathcal{L}_\mathrm{ghost} = 
\left( \Box c_1 \right) b_0
+ \left( - c_1 + \Box c_2 \right) b_1 + \cdots 
+ \left( - c_{n-1} + \Box c_n \right) b_{n-1}  
 - c_n b_n \, .
\end{align}
According to this formalism,  the action (\ref{hfr2}) can be rewritten as 
\begin{equation}
\label{hfr12}
S = \frac{1}{2\kappa^2} \int d^4 x \sqrt{-g} \left\{ f \left( \phi_0, \phi_1, \cdots , \phi_n \right) 
+ \e^{-\rho } R + \mathcal{L}_\mathrm{GF} \right\} \, .
\end{equation}
Then instead of the action  (\ref{hfr2}), if we consider the following action,  
\begin{equation}
\label{hfr13}
S = \frac{1}{2\kappa^2} \int d^4 x \sqrt{-g} \left\{ f \left( \phi_0, \phi_1, \cdots , \phi_n \right) 
+ \e^{-\rho } R + \mathcal{L}_\mathrm{GF} + \mathcal{L}_\mathrm{ghost} \right\} \, ,
\end{equation}
the action (\ref{hfr13}) is invariant under the BRS transformation (\ref{hfr10}). 
The corresponding BRS current is given by 
\begin{equation}
\label{BRS1}
J_\mathrm{BRS}^\mu = - \frac{\e^{\rho}}{\kappa^2}\sum_{i=1}^n c_i \partial^\mu \phi_{i-1}\, ,
\end{equation}
and the BRS charge $Q_\mathrm{BRS}$ is defined as 
\begin{equation}
\label{BRS2}
Q_\mathrm{BRS} = \int d\Sigma_\mu \sqrt{\gamma} J_\mathrm{BRS}^\mu \, .
\end{equation}
Here $\Sigma_\mu$ expresses the space-like surface and $\gamma$ is the metric induced 
on $\Sigma_\mu$. 
Then in quantum theory, by imposing the condition that the physical states should be 
annihilated by the BRS charge $Q_\mathrm{BRS}$, the quantum fluctuations of $\phi_i$ 
and $c_i$ are prohibited. Although the quantum fluctuations of $\lambda_i$ and $b_i$ are not 
prohibited, the fluctuations only generate the zero-norm states and therefore the 
fluctuations does not contribute to any physical process. Therefore ghost can be eliminated 
(see also \cite{Biswas:2005qr}).

As long as we consider the classical background, where the ghosts $c_i$ and the 
anti-ghosts $b_i$ vanish, i.e. $c_i=b_i=0$, the action (\ref{hfr13}) is equivalent to 
(\ref{hfr2}) and therefore (\ref{hfr1}). 
The condition $c_i=b_i=0$ is also required by the super-selection rule or ghost number 
conservation. 

%%%%%%%%%%%%%%%%%%%%%%%%%%%%%%%%%%%%%

Let us first  consider the classical solution which satisfies Eqs.~(\ref{hfr5}), (\ref{hfr6}), 
(\ref{hfr7}), (\ref{hfr8}), and (\ref{hfr9}).  
We assume that all the scalar fields $\phi_0$, $\phi_1$, $\cdots$, $\phi_n$, $\rho$, 
$\lambda_1$, $\cdots$, $\lambda_n$ are constant. 
Then we find 
\begin{align}
\label{hfr15}
0 =&  - 2 f \left( \phi_0, \phi_1, \cdots , \phi_n \right) + \e^{-\rho } \phi_0 \, , \\
\label{hfr16}
0 =& \e^{2\rho} \phi_i \, , \quad i=1,2,\cdots, n \, ,\\
\label{hfr17}
0=& \e^{2\rho } \frac{\partial f \left( \phi_0, \phi_1, \cdots , \phi_n \right) }{\partial \phi_0} 
 - \e^{\rho } \, , \\
\label{hfr18}
0=& \frac{\partial f \left( \phi_0, \phi_1, \cdots , \phi_n \right) }{\partial \phi_i} 
+ \lambda_i \, , \quad i=1,2,\cdots, n \, .
\end{align}
Eq.~(\ref{hfr16}) tells $\phi_i=0$, for $i=1,2,\cdots, n$. 
As an example, we may consider \cite{CCC}, 
\begin{equation}
\label{hfr14}
 f \left( R, \Box R, \Box^2 R, \cdots, \Box^n R \right) 
= R + \sum_{i=1}^n \alpha_i R \Box^i R \, .
\end{equation}
Here $\alpha_i$'s are dimensional coupling constants. 
From Eq.~(\ref{hfr14}), Eqs.~(\ref{hfr15}), (\ref{hfr17}), and (\ref{hfr18}) have the following form, 
\begin{align}
\label{hfr20}
0 =&  - 2 \phi_0 + \e^{-\rho } \phi_0 \, , \\
\label{hfr22}
0=& \e^{2\rho } - \e^{\rho } \, , \\
\label{hfr23}
0=& \alpha_i \phi_0 + \lambda_i \, , \quad i=1,2,\cdots, n \, .
\end{align}
Then the solution is given by 
\begin{equation}
\label{hfr24}
\rho=\phi_0 = \phi_i = \lambda_i = 0 \, , \quad i=1,2,\cdots, n \, ,
\end{equation}
which corresponds to the flat space-time. 
Then when we consider the perturbation from the solution (\ref{hfr24}) in the action 
(\ref{hfr4}) with (\ref{hfr14}), we find 
\begin{align}
\label{hfr25}
& S = \frac{1}{2\kappa^2} \int d^4 x \sqrt{-g} \left\{ R 
 - \frac{3}{2} \partial_\mu \rho \partial^\mu \rho 
 - \partial_\mu \lambda_1 \partial^\mu \phi_0 
 - \partial_\mu \lambda_2 \partial^\mu \phi_1 + \cdots 
 - \partial_\mu \lambda_n \partial^\mu \phi_{n-1} \right. \nn
& \qquad \qquad \qquad \qquad \left. 
+  \phi_0 \sum_{i=1}^n \alpha_i \phi_i + 3 \rho\phi_0 - \lambda_1 \phi_1 
 - \lambda_2 \phi_2 - \cdots - \lambda_n \phi_n \right\} \, .
\end{align}
It is worth noticing  that the case  (\ref{hfr14}) and therefore the action (\ref{hfr25}) 
are linear with respect to $\phi_n$. Here Eq.~(\ref{hfr9}) gives a constraint. 
In fact, the variation with respect to $\phi_n$ gives $\lambda_n = \alpha_n \phi_0$. 
Then deleting $\lambda_n$, we obtain 
\begin{align}
\label{hfr26}
& S = \frac{1}{2\kappa^2} \int d^4 x \sqrt{-g} \left\{ R 
 - \frac{3}{2} \partial_\mu \rho \partial^\mu \rho 
 - \partial_\mu \lambda_1 \partial^\mu \phi_0 
 - \partial_\mu \lambda_2 \partial^\mu \phi_1 + \cdots 
 - \alpha_n \partial_\mu \phi_0 \partial^\mu \phi_{n-1} \right. \nn
& \qquad \qquad \qquad \qquad \left. 
+ \phi_0 \sum_{i=1}^{n-1} \alpha_i \phi_i + 3 \rho\phi_0 - \lambda_1 \phi_1 
 - \lambda_2 \phi_2 - \cdots - \lambda_{n-1} \phi_{n-1} \right\} \, .
\end{align}

First we consider the simplest case where $n=1$. It is
\begin{equation}
\label{hfr27}
S = \frac{1}{2\kappa^2} \int d^4 x \sqrt{-g} \left\{ R 
 - \frac{3}{2} \partial_\mu \rho \partial^\mu \rho 
 - \alpha_1 \partial_\mu \phi_0 \partial^\mu \phi_0 + 3 \rho\phi_0 \right\} \, .
\end{equation}
We  note that, for  $\alpha_1>0$, there is no ghost. 
If we redefine the fields $\rho$ and $\phi_0$ by the new fields $\xi_\pm$ as, 
\begin{equation}
\label{hfr28}
\rho = \frac{1}{\sqrt{6}} \left( \xi_+ + \xi_- \right)\, , \quad 
\phi_0 = \frac{1}{\sqrt{2\alpha_1}}  \left( \xi_+ - \xi_- \right)\, , 
\end{equation}
the action (\ref{hfr27}) can be rewritten as 
\begin{equation}
\label{hfr29}
S = \frac{1}{2\kappa^2} \int d^4 x \sqrt{-g} \left\{ R 
 - \frac{1}{2} \partial_\mu \xi_+ \partial^\mu \xi_+ 
 - \frac{1}{2} \partial_\mu \xi_- \partial^\mu \xi_- 
 - \frac{1}{2}\sqrt{\frac{3}{\alpha_1}} \left( - \xi_+^2 + \xi_-^2 \right) \right\} \, .
\end{equation}
Therefore $\xi_+$ becomes a tachyon. 

We now consider the case $n=2$, 
%% diminesions $\phi_0:2$, $\lambda_1:-2$, $\phi_1:4$, $\alpha_1: -4$, $\alpha_2:-6$ 
\begin{equation}
\label{hfr30}
S = \frac{1}{2\kappa^2} \int d^4 x \sqrt{-g} \left\{ R 
 - \frac{3}{2} \partial_\mu \rho \partial^\mu \rho 
 - \partial_\mu \lambda_1 \partial^\mu \phi_0 
 - \alpha_2 \partial_\mu \phi_0 \partial^\mu \phi_1 
+ \alpha_1 \phi_0 \phi_1 + 3 \rho\phi_0 - \lambda_1 \phi_1 \right\} \, .
\end{equation}
The variations with respect to $\lambda_1$ and $\phi_1$ give
\begin{equation}
\label{hfr31}
0 = \Box \phi_0 - \phi_1 \, , \quad 
0 = \alpha_2 \Box \phi_0 + \alpha_1 \phi_0 - \lambda_1 \, .
\end{equation}
Then by combining these equations into (\ref{hfr31}), we get
\begin{equation}
\label{hfr32}
\lambda_1 = \alpha_1 \phi_0 + \alpha_2 \phi_1 \, .
\end{equation}
We can delete $\lambda_1$ in the action (\ref{hfr30}) and obtain
\begin{align}
\label{hfr33}
S =& \frac{1}{2\kappa^2} \int d^4 x \sqrt{-g} \left\{ R 
 - \frac{3}{2} \partial_\mu \rho \partial^\mu \rho 
 - \alpha_1 \partial_\mu \phi_0 \partial^\mu \phi_0 
 - 2 \alpha_2 \partial_\mu \phi_0 \partial^\mu \phi_1 
+ 3 \rho\phi_0 - \alpha_2 \phi_1^2 \right\} \nn
=& \frac{1}{2\kappa^2} \int d^4 x \sqrt{-g} \left\{ R 
 - \frac{3}{2} \partial_\mu \rho \partial^\mu \rho 
 - \alpha_1 \partial_\mu {\tilde\phi}_0 \partial^\mu \tilde\phi_0 
+ \frac{\alpha_2^2}{\alpha_1} \partial_\mu \phi_1 \partial^\mu \phi_1 
+ 3 \rho {\tilde\phi}_0 
 - 3 \frac{\alpha_2}{\alpha_1} \rho \phi_1 - \alpha_2 \phi_1^2 \right\} \, , \nn
{\tilde\phi}_0 \equiv& \phi_0 + \frac{\alpha_2}{\alpha_1} \phi_1
\, .
\end{align}
If $\alpha_1>0$, ${\tilde\phi}_0$ becomes canonical but $\phi_1$ becomes a ghost. 
By redefining the scalar fields,  
\begin{equation}
\label{diag1}
\rho=\frac{1}{\sqrt{6}} \left( \xi + \zeta \right) \, , \quad 
{\tilde\phi}_0 = \frac{1}{2\sqrt{\alpha_1}}\left( \xi + \zeta \right) \, ,\quad
\phi_1 = \frac{1}{\alpha_2} \sqrt{\frac{\alpha_1}{2}}\eta\, ,
\end{equation}
%%%%%%%% ?½È‰ï¿½?½C?½?½
we obtain, 
\begin{equation}
\label{diag2}
S= \frac{1}{2\kappa^2} \int d^4 x \sqrt{-g} \left\{ R 
 - \frac{1}{2} \partial_\mu \xi \partial^\mu \xi 
 - \frac{1}{2} \partial_\mu \zeta \partial^\mu \zeta 
+ \frac{1}{2} \partial_\mu \eta \partial^\mu \eta 
+ \frac{1}{2}\sqrt{\frac{3}{2\alpha_1}}\left( \xi^2 - \zeta^2 \right) 
 - \frac{1}{2}\sqrt{\frac{3}{\alpha_1}} \left( \xi + \zeta \right) \eta 
 - \frac{\alpha_1}{2 \alpha_2} \eta^2 \right\} \ . 
\end{equation}
The spectrum of the scalar modes can be found by investigating the following $3\times 3$ 
matrix in the momentum space.
\begin{equation}
\label{diag3}
\left( M_{ij} \right) = \left( \begin{array}{ccc}
 -k^2 + \sqrt{\frac{3}{2\alpha_1}} & 0 & - \frac{1}{2}\sqrt{\frac{3}{2\alpha_1}} \\
0 & - k^2 - \sqrt{\frac{3}{2\alpha_1}} & - \frac{1}{2}\sqrt{\frac{3}{2\alpha_1}} \\
 - \frac{1}{2}\sqrt{\frac{3}{2\alpha_1}} & 
 - \frac{1}{2}\sqrt{\frac{3}{2\alpha_1}} & k^2 - \frac{\alpha_1}{\alpha_2} 
\end{array} \right) 
= \left( \begin{array}{ccc}
 -k^2 + A & 0 & B \\
0 & - k^2 - A & B \\
B & B & k^2 - C
\end{array} \right) \, .
\end{equation}
Here
\begin{equation}
\label{diag3b}
A \equiv \sqrt{\frac{3}{2\alpha_1}} \, , \quad 
B \equiv - \frac{1}{2}\sqrt{\frac{3}{2\alpha_1}} \, , \quad 
C \equiv \frac{\alpha_1}{\alpha_2} \, .
\end{equation}
In order to investigate the position of the poles in the propagator, which correspond to 
the mass of the scalar fields, we consider the determinant of the matrix $M$, 
\begin{align}
\label{diag4}
0 = \det M =& \left( k^4 - A^2 \right) \left(k^2 - C \right) + 2 B^2 k^2 \nn
=& \left( k^2 - \frac{C}{3} \right)^3 + \left( - \frac{C^2}{3} - A^2 + 2B^2 \right) 
\left( k^2 - \frac{C}{3} \right)
 - \frac{2C^3}{27} + \frac{2 A^2 C}{3} + \frac{2 B^2 C}{3} \, .
\end{align}
By assuming 
\begin{align}
\label{diag5}
\det M =& \left( k^2 - \frac{C}{3} - \alpha - \beta \right) \left( k^2 - \frac{C}{3} - \alpha \xi - \beta \xi^2 \right) 
\left( k^2 - \frac{C}{3} - \alpha \xi^2 - \beta \xi \right) \nn
=& \left( k^2 - \frac{C}{3} \right)^3 - 3 \alpha \beta \left( k^2 - \frac{C}{3} \right) - \alpha^3 - \beta^3 \, ,
\end{align}
with $\xi \equiv \e^{i\frac{\pi}{3}}$, we find 
\begin{equation}
\label{diag6}
 - 3\alpha \beta = - \frac{C^2}{3} - A^2 + 2B^2 \, , \quad 
\alpha^3 + \beta^3 = - \left(  - \frac{2C^3}{27} + \frac{2 A^2 C}{3} + \frac{2 B^2 C}{3} \right) \, .
\end{equation}
Therefore we obtain 
\begin{equation}
\label{diag7}
0 = \left( \alpha^3 \right)^2 + \frac{2C}{3} \left( - \frac{C^2}{3} + A^2 + B^2 \right) \alpha^3 
 - \frac{1}{27} \left( - \frac{C^2}{3} - A^2 + 2B^2 \right)^3 \, ,
\end{equation}
and 
\begin{equation}
\label{diag8}
\alpha^3 = \gamma_\pm \equiv 
\frac{C}{3} \left( - \frac{C^2}{3} + A^2 + B^2 \right) 
\pm \sqrt{ \frac{C^2}{9} \left( - \frac{C^2}{3} + A^2 + B^2 \right)^2 
+ \frac{1}{27} \left( - \frac{C^2}{3} - A^2 + 2B^2 \right)^3 } \, , 
\quad \beta^3 = \gamma_\mp
\end{equation}
or
\begin{equation}
\label{diag9}
\gamma_\pm = 
\frac{\alpha_1}{3\alpha_2} \left( - \frac{\alpha_1^2}{3 \alpha_2^2} 
+ \frac{15}{8\alpha_1} \right) 
\pm \sqrt{ \frac{\alpha_1^2}{9\alpha_2^2} \left( - \frac{\alpha_1^2}{3 \alpha_2^2} 
+ \frac{15}{8\alpha_1} \right)^2
+ \frac{1}{27} \left( - \frac{\alpha_1^2}{3 \alpha_2^2} - \frac{3}{4\alpha_1} \right)^3 } \, .
\end{equation}
If 
\begin{equation}
\label{diag10}
D \equiv \frac{\alpha_1^2}{9\alpha_2^2} \left( - \frac{\alpha_1^2}{3 \alpha_2^2} 
+ \frac{15}{8\alpha_1} \right)^2
+ \frac{1}{27} \left( - \frac{\alpha_1^2}{3 \alpha_2^2} - \frac{3}{4\alpha_1} \right)^3
> 0 \, ,
\end{equation}
$\gamma_+$ and $\gamma_-$ are real numbers and therefore 
$\alpha\xi + \beta\xi^2$ and $\alpha\xi^2 + \beta\xi$ are complex numbers 
although $\alpha + \beta$ is a real number. 
Therefore if $D>0$, there appear one real mass squared and two complex masses. 
On the other hand, if $D<0$, $\gamma_-$ is a complex conjugate of $\gamma_+$, 
that is, $\gamma_-= \left( \gamma_+ \right)^\dagger$ and therefore $\beta=\alpha^\dagger$. 
Then all of $\alpha\xi + \beta\xi^2$, $\alpha\xi^2 + \beta\xi$, and 
$\alpha + \beta$ are real numbers.  
Therefore if $D<0$, all of the mass squared are real numbers although they might 
be negative, that is, the corresponding modes might be tachyon. 

As an example of non-linear action (\ref{hfr25})  with respect to $\phi_n$, we consider 
the model with $n=1$, 
\begin{equation}
\label{hfr34}
 f \left( R, \Box R \right) 
= R + \alpha \left(\Box R\right)^2 \, ,
\end{equation}
with a constant $\alpha$. 
Then Eqs.~(\ref{hfr15}), (\ref{hfr16}), (\ref{hfr17}), and (\ref{hfr18}) have the following forms, 
\begin{align}
\label{hfr35}
0 =&  - 2 \left( \phi_0 + \alpha \phi_1^2 \right) + \e^{-\rho } \phi_0 \, , \\
\label{hfr36}
0 =& \e^{2\rho} \phi_1\, ,\\
\label{hfr37}
0=& \e^{2\rho } - \e^{\rho } \, , \\
\label{hfr18B}
0=& 2\alpha \phi_1 + \lambda_1 \, .
\end{align}
The background solution is given by,
\begin{equation}
\label{hfr38}
\phi_0 = \phi_1 = \lambda_1 = \rho = 0 \, .
\end{equation}
Then the perturbation from the solution (\ref{hfr38}) in the action 
(\ref{hfr4}) with (\ref{hfr34}) gives,
\begin{equation}
\label{hfr39}
S = \frac{1}{2\kappa^2} \int d^4 x \sqrt{-g} \left\{ R 
 - \frac{3}{2} \partial_\mu \rho \partial^\mu \rho 
 - \partial_\mu \lambda_1 \partial^\mu \phi_0 
+ \rho\phi_0 + \alpha \phi_1^2 + \lambda_1 \phi_1 \right\}\, .
\end{equation}
The variation with respect to $\phi_1$ gives, 
\begin{equation}
\label{hfr40}
\lambda_1 = - 2\alpha \phi_1\, .
\end{equation}
By deleting $\lambda_1$ in the action (\ref{hfr39}), we obtain
\begin{equation}
\label{hfr41}
S = \frac{1}{2\kappa^2} \int d^4 x \sqrt{-g} \left\{ R 
 - \frac{3}{2} \partial_\mu \rho \partial^\mu \rho 
+ 2 \alpha \partial_\mu \phi_1 \partial^\mu \phi_0 
+ \rho\phi_0 - \alpha \phi_1^2 \right\}\, .
\end{equation}
If we define the new fields $\zeta_\pm$ by 
\begin{equation}
\label{hfr42}
\phi_1 = \frac{1}{2\alpha} \left( \zeta_+ - \zeta_- \right) \, , \quad 
\phi_0 = \frac{1}{2} \left( \zeta_+ - \zeta_- \right) \, , 
\end{equation}
we can further rewrite the action (\ref{hfr41}) as follows,  
\begin{equation}
\label{hfr43}
S = \frac{1}{2\kappa^2} \int d^4 x \sqrt{-g} \left\{ R 
 - \frac{3}{2} \partial_\mu \rho \partial^\mu \rho 
+ \frac{1}{2} \partial_\mu \zeta_+ \partial^\mu \zeta_+  
 - \frac{1}{2} \partial_\mu \zeta_- \partial^\mu \zeta_-  
+ \frac{1}{2} \left( \zeta_+ - \zeta_- \right) \rho 
 - \frac{1}{4\alpha} \left( \zeta_+ - \zeta_- \right)^2 \right\}\, .
\end{equation}
Therefore $\zeta_+$ is a ghost field. 

Similar considerations can be developed also in higher-order non-local case. Starting from these results the theory of  GWs can be developed.

\section{Higher-order non-local gravity}\label{3}

Almost a decade ago, a non-local modification of the Einstein-Hilbert (EH) action has been proposed \cite{Deser:2007jk}, and the new action has the following form
 \begin{eqnarray}
 	\mathcal{S}_{\rm standard-NL}&=&\frac{1}{2\kappa^2}\int d^{4}x\, \sqrt{-g(x)} \, R(x)\left[1 + f\Big((\square^{-1}R)(x)\Big) \right]+\int d^{4}x\, \sqrt{-g(x)}\,{\cal L}_{m}\,, \label{1b}
 \end{eqnarray}	 
 where, as above,  $\kappa=8\pi$, $G=c=1$, $R$ is the Ricci scalar,  $f$ is an arbitrary function which depends on the retarded Green function evaluated at the Ricci scalar, ${\cal L}_{m}$ is any matter Lagrangian and  $\square \equiv \partial_{\rho}(e g^{\sigma\rho}\partial_{\sigma})/e$ is the above d'Alembert
 scalar-wave operator, which can be written in terms of the  Green function $G(x,x')$ as 
 \begin{eqnarray}
 	(\square^{-1}F)(x)=\int d^4x'\, e(x') F(x')G(x,x')\,.\label{G}
 \end{eqnarray}
 It is clear that by setting $f(\square^{-1}R)=0$, the above action is equivalent to the Einstein-Hilbert one plus the matter content.  The non-locality is introduced by the inverse of the d'Alembert operator. Corrections of this kind arise naturally as soon as quantum loop effects are studied and they are also considered as possible solution to the black hole information paradox \cite{Donoghue:1994dn,Giddings:2006sj}. Since then, a lot of studies of non-localities have been done in various contexts \cite{univ4,modesto1,modesto2,st1,st2,loop,jm}. 
 In Refs. \cite{Mod1,Mod3,Mod4,Mod5}, non-local quantum gravity is fully discussed putting in evidence results and open issues.  From the string  theory point of view, in \cite{Arefeva:2007wvo},   some bouncing solutions are presented.  In \cite{Arefeva:2007xdy} solutions of an expanding Universe with phantom dark energy are reported while,  in \cite{Barnaby:2008fk},   non-Gaussianities during inflation are discussed. Starting from IR scales, a lot of progress has also been done. Unification of inflation with late-time acceleration, as well as, the dynamics of a local form of the theory have been studied in \cite{Nojiri:2007uq, Jhingan:2008ym}. In \cite{Deser:2013uya}, they show that non-local gravity models do not alter the GR predictions for gravitationally bound systems, and also they are ghost-free and stable. Finally, in \cite{Deffayet:2009ca,Koivisto:2008dh,Koivisto:2008xfa}, they derived a technique to fix the functional form of  $f$ in the action, which is called non-local distortion function. For details see \cite{Barvinsky:2014lja}, where  non-local aspects both from the quantum-field theory point of view and from the cosmological one are summarized.

In order to start our considerations, let us define the following non-local action 
\begin{equation}
\label{nlhfr1}
S = \frac{1}{2\kappa^2} 
\int d^4 x \sqrt{-g} f \left( R, \Box^{-1} R, \Box^{-2} R, \cdots, \Box^{-n} R \right) \, .
\end{equation}
As above, by introducing the scalar fields $\phi_0$, $\phi_1$, $\cdots$ $\phi_n$, $\rho $, and the Lagrange multipliers $\lambda_1$, 
$\cdots$, $\lambda_n$, the action (\ref{hfr1}) can be rewritten as follows,
\begin{align}
\label{nlhfr2}
S =& \frac{1}{2\kappa^2} \int d^4 x \sqrt{-g} \left\{ f \left( \phi_0, \phi_1, \cdots , \phi_n \right) 
+ \e^{-\rho } \left( R - \phi_0 \right) + \lambda_1 \left( \phi_0 - \Box \phi_1 \right) 
+ \lambda_2 \left( \phi_1 - \Box \phi_2 \right) \right. \nn
& \left. + \cdots 
+ \lambda_n \left( \phi_{n-1} - \Box \phi_n \right) 
\right\} \, .
\end{align}
Rescaling by the metric (\ref{hfr3}),  we can rewrite 
the action (\ref{nlhfr2})  as
\begin{align}
\label{nlhfr4}
& S = \frac{1}{2\kappa^2} \int d^4 x \sqrt{-g} \Bigl\{ R 
 - \frac{3}{2} \partial_\mu \rho \partial^\mu \rho 
+ \e^{\rho } \left( \partial_\mu \lambda_1 \partial^\mu \phi_1 
+ \partial_\mu \lambda_2 \partial^\mu \phi_2 + \cdots 
+ \partial_\mu \lambda_n \partial^\mu \phi_n \right) \nn
& \qquad \qquad \qquad \qquad 
 - V \left( \phi_0, \phi_1, \cdots \phi_n, \rho , \lambda_1, \cdots \lambda_n \right) 
\Bigr\} \, , \nn
& V \left( \phi_0, \phi_1, \cdots \phi_n, \rho , \lambda_1, \cdots \lambda_n \right) 
\equiv \e^{2\rho } \left( - f \left( \phi_0, \phi_1, \cdots , \phi_n \right) 
+ \e^{-\rho } \phi_0 - \lambda_1 \phi_0 - \lambda_2 \phi_1 + \cdots 
+ \lambda_n \phi_{n-1} \right) \, .
\end{align}
Then the field equations are given by
\begin{align}
\label{nlhfr5}
0 =& 3 \Box \rho 
 - \e^{\rho } \left( \partial_\mu \lambda_1 \partial^\mu \phi_1 
+ \partial_\mu \lambda_2 \partial^\mu \phi_2 + \cdots 
+ \partial_\mu \lambda_n \partial^\mu \phi_n \right) 
 - \e^{2\rho } \left( - 2 f \left( \phi_0, \phi_1, \cdots , \phi_n \right) 
+ \e^{-\rho } \phi_0 \right) \, , \\
\label{nlhfr6}
0 =& - \nabla_\mu \left( \e^{\rho } \partial^\mu \phi_i \right) 
 - \e^{2\rho } \phi_{i-1} \, , \quad i=1,2,\cdots,n\, , \\
\label{nlhfr7}
0=& - \nabla_\mu \left( \e^{\rho } \partial^\mu \lambda_n \right) 
+ \e^{2\rho } \frac{\partial f \left( \phi_0, \phi_1, \cdots , \phi_n \right) }{\partial \phi_n} \, , \\
\label{nlhfr8}
0=& - \nabla_\mu \left( \e^{\rho } \partial^\mu \lambda_i \right)
+ \e^{2\rho } \left( 
\frac{\partial f \left( \phi_0, \phi_1, \cdots , \phi_n \right) }{\partial \phi_i} 
+ \lambda_{i+1} \right)\, , \quad i=1,2,\cdots,n-1 \, , \\
\label{nlhfr9}
0=& \e^{2\rho } \left( 
\frac{\partial f \left( \phi_0, \phi_1, \cdots , \phi_n \right) }{\partial \phi_0} 
 - \e^{-\rho} \right) \, .
\end{align}
Because the action (\ref{nlhfr4}) does not include the derivative term of $\phi_0$, 
the scalar field $\phi_0$ is just a non-dynamical auxiliary field. 
Therefore besides the metric, the dynamical fields are $\rho$, $\lambda_i$, and 
$\phi_i$, where $i=1,2,\cdots,n$. 
In $\lambda_i$ and $\phi_i$ $\left(i=1,2,\cdots,n\right)$, half of them are canonical 
but the remaining half are ghosts.

Under the above BRS transformation (\ref{hfr10}), we find 
\begin{align}
\label{nlhfr11}
& \delta \left( \left( - \e^{-\rho_0} + \lambda_1 \right) b_0
+ \left( - \Box \lambda_1 + \lambda_2 \right) b_1 + \cdots 
+ \left( - \Box \lambda_{n-1} + \lambda_n \right) b_{n-1}  
 - \Box \lambda_n b_n \right) =  \epsilon \left( \mathcal{L}_\mathrm{GF} 
+ \mathcal{L}_\mathrm{ghost} \right) \, , \nn
& \mathcal{L}_\mathrm{GF} = \left( - \e^{-\rho_0} + \lambda_1 \right) \phi_0
+ \left( - \Box \lambda_1 + \lambda_2 \right) \phi_1 + \cdots 
+ \left( - \Box \lambda_{n-1} + \lambda_n \right) \phi_{n-1}  
 - \Box \lambda_n \phi_n \, , \nn
& \mathcal{L}_\mathrm{ghost} = c_1 b_0 + \left( - \Box c_1 + c_2 \right) b_1 + \cdots 
+ \left( - \Box c_{n-1} + c_n \right) b_{n-1} - \Box c_n b_n \, .
\end{align}
Then, as in (\ref{hfr12}), the action (\ref{nlhfr2}) can be rewritten as 
\begin{equation}
\label{nlhfr12}
S = \frac{1}{2\kappa^2} \int d^4 x \sqrt{-g} \left\{ f \left( \phi_0, \phi_1, \cdots , \phi_n \right) 
+ \e^{-\rho } R + \mathcal{L}_\mathrm{GF} \right\} \, .
\end{equation}
Instead of  action  (\ref{nlhfr2}), if we consider the following action,  
\begin{equation}
\label{nlhfr13NL}
S = \frac{1}{2\kappa^2} \int d^4 x \sqrt{-g} \left\{ f \left( \phi_0, \phi_1, \cdots , \phi_n \right) 
+ \e^{-\rho } R + \mathcal{L}_\mathrm{GF} + \mathcal{L}_\mathrm{ghost} \right\} \, ,
\end{equation}
the action (\ref{nlhfr13NL}) is invariant under the BRS transformation (\ref{hfr10}), again. 
As in Eq.(\ref{BRS1}), the BRS current is given by 
\begin{equation}
\label{BRS3}
J_\mathrm{BRS}^\mu = - \frac{\e^{\rho}}{\kappa^2}\sum_{i=1}^n c_i \partial^\mu \phi_i\, .
\end{equation}
Then by defining the BRS charge $Q_\mathrm{BRS}$ as in (\ref{BRS2}) and 
the physical states which are annihilated by, we may be able to eliminate the ghosts 
(see also \cite{Conroy:2014eja} for the $R f \left( \Box^{-1} \right) R$ gravity case). 

The action (\ref{nlhfr13NL}) tells that the physical spectrum is formally not changed from that in 
the action (\ref{hfr13}). This result will be very important in order to develop  GWs considerations  for non-local gravity.

\section{Considerations on Gravitational waves in local and non-local gravity}\label{4}
To analyze GWs in non-local gravity, we will use the formal equivalence between the spectrum of the actions \eqref{hfr1} and \eqref{nlhfr1} demonstrated in the previous section. In particular we derive the properties of the gravitational radiation in the non-local model, linear with respect to both $R$ and $\Box^{-i}R$, that is 
\begin{equation}
\label{nlhfr14}
 f \left( R, \Box^{-1} R, \Box^{-2} R, \cdots, \Box^{-n} R \right) 
= R + \sum_{i=1}^n \alpha_i R \Box^{-i} R \, ,
\end{equation}
analogue to the model  \eqref{hfr14}, linear with respect to both $R$ and $\Box^i R$. 

Let us start  with linearizing    higher order gravity  \eqref{hfr14}, and  expand the metric tensor $g_{\mu\nu}$  at first order  in $h_{\mu\nu}$ with respect to the flat metric $\eta_{\mu\nu}$ 
\begin{equation}
g_{\mu\nu}=\eta_{\mu\nu}+h_{\mu\nu}+\mathcal{O}\left(h^2\right)\ .
\end{equation}
We get the linearized field equation in matter and in harmonic gauge \cite{CCC,CCC1}
\begin{equation}
\label{hfr44}
\Box\bar{h}_{\mu\nu}+2\sum_{k=1}^{n}\alpha_{k}\left(\eta_{\mu\nu}\Box^{k+2}-\partial_{\mu}\partial_{\nu}\Box^{k+1}\right)\bar{h}=-2\kappa^{2}\mathcal{T}_{\mu\nu}^{(0)}
\ ,
\end{equation}
and its trace is 
\begin{equation}
\label{hfr45}
\Box\bar{h}-6\sum_{k=1}^{n}\alpha_{k}\Box^{k+2}\bar{h}=-2\kappa^{2}\mathcal{T}^{\left(0\right)}\ ,
\end{equation}
where $\mathcal{T}_{\mu\nu}^{(0)}$ is the zero-order energy-momentum tensor of matter and $\bar{h}_{\mu\nu}$ is set to 
\begin{equation}
\bar{h}_{\mu\nu}=h_{\mu\nu}-\frac{1}{2}\eta_{\mu\nu}h\ .
\end{equation}
Now we solve the linear homogeneous PDEs associated with Eqs \eqref{hfr44} and \eqref{hfr45}  adopting a distribution calculus framework, because $\bar{h}\left(x\right)\in\mathcal{S}'(\mathbb{R}^{4})$ is a continuous linear functional  of the space of all tempered distributions on $\mathbb{R}^{4}$ closed under Fourier transformation, or the dual of the Schwartz  space $\mathcal{S}(\mathbb{R}^{4})$ of rapidly decreasing functions. See  \cite{CCC} for details.
  
 In $k$-space the trace equation \eqref{hfr45} in vacuum  becomes
\begin{equation}
\label{Tracecekspace}
\left(k^2+6\sum_{l=1}^{n}\alpha_{l}\left(-1\right)^{l}k^{2(l+2)}\right)\hat{h}\left(k\right)=0\ ,
\end{equation}
where $\hat{h}\left(k\right)$ is the Fourier transformation of the metric perturbation $h(x)$.
By the mean of properties of $\delta$-distribution we obtain
\begin{equation}\label{AkB}
\hat{h}\left(k\right)=\sqrt{2\pi}\sum_{m=1}^{n+2}\left[\delta\left(k^{0}-\omega_{m}\right)\hat{B}_{m}\left(\mathbf{k}\right)+\delta\left(k^{0}+\omega_{m}\right)\hat{B}_{m}^{*}\left(-\mathbf{k}\right)\right]\ ,
\end{equation}
with 
\begin{equation}
\hat{B}_{m}\left(\mathbf{k}\right)=\frac{Q_{m}\left(\mathbf{k}\right)}{2\sqrt{2\pi}\omega_{m}\vert 6\sum_{l=1}^{n}\left(l+2\right)\alpha_{l}\left(-1\right)^{l}\omega_{m}^{2\left(l+1\right)}+1\vert}\ .
\end{equation}
where $Q_{m}\left(\mathbf{k}\right)$ is a suitable complex function, while $\omega_{m}$ are the $m$-frequencies 
\begin{equation}
\omega_{m}=\sqrt{M_{m}^{2}+\vert\mathbf{k}\vert^{2}}\ ,
\end{equation}
with $M_{m}^{2}$ the $n+2$ solutions of the linear equation in $k^2$
\begin{equation}
k^2+6\sum_{l=1}^{n}\alpha_{l}\left(-1\right)^{l}k^{2(l+2)}=0\ ,
\end{equation}
under suitable hypothesis for the coefficients $\alpha_{l}$. We now perform the inverse Fourier transform of $\hat{h}\left(k\right)$ as
\begin{equation}
\bar{h}\left(x\right)=\int \frac{d^{4}k}{\left(2\pi\right)^{2}}\hat{h}\left(k\right)e^{ik^{\alpha}x_{\alpha}}\ ,
\end{equation}
and from Eq.\eqref{hfr44} we derive  the gravitation waves in vacuum in higher order gravity, linear with respect to both $R$ and $\Box^i R$ that is \cite{CCC}
\begin{align}\label{GWEFTBTG}
h_{\mu\nu}\left(x\right)&=\int \frac{d^{3}\mathbf{k}}{(2\pi)^{3/2}}C_{\mu\nu}\left(\mathbf{k}\right)e^{ik_{1}^{\alpha}x_{\alpha}} \nonumber\\
&+\sum_{m=2}^{n+2}\int \frac{d^{3}\mathbf{k}}{(2\pi)^{3/2}}\left\{\frac{1}{3}\left[\frac{\eta_{\mu\nu}}{2}+\frac{\left(k_{m}\right)_{\mu}\left(k_{m}\right)_{\nu}}{k_{m}^{2}}\right]\right\}\hat{B}_{m}\left(\mathbf{k}\right)e^{ik_{m}^{\alpha}x_{\alpha}}+c.c.\ .
\end{align}
To study polarizations and oscillation modes of waves \eqref{GWEFTBTG} we use the equation of geodesic deviation for a wave travelling along $+z$-direction in a local proper reference frame 
\begin{equation}\label{eqdevgeoelectric}
\ddot x^{i}=-R^{i}_{\phantom{i}0k0}x^{k}\ ,
\end{equation}
where a deviation vector is $\left(x^1,x^2,x^3\right)$, the latin index range over the set $\left\{1,2,3\right\}$ and $R^{i}_{\phantom{i}0k0}$ are so-called "electric" components of the Riemann tensor which is expressed in terms of linear perturbation $h_{\mu\nu}$ as 
\begin{equation}\label{relaxionRh}
R^{\left(1\right)}_{\phantom{1}i0j0}=\frac{1}{2}\left(h_{i0,0j}+h_{0j,i0}-h_{ij,00}-h_{00,ij}\right)\ .
\end{equation}
From Eqs. \eqref{eqdevgeoelectric} and \eqref{relaxionRh} we have 
\begin{equation}\label{eqdevgeolinear}
\begin{cases}
\ddot x(t)=-\frac{1}{2}\left(xh_{11,00}+yh_{12,00}\right) \\
\ddot y(t)=-\frac{1}{2}\left(xh_{12,00}+yh_{11,00}\right)\\
\ddot z(t)=\frac{1}{2}\left(2h_{03,03}-h_{33,00}-h_{00,33}\right)z
\end{cases}\ .
\end{equation}
Keeping $\mathbf{k}$ fixed,  for a plane wave with $k_{1}^{2}=0$  where $k_{1}^{\mu}=\left(\omega_{1},0,0,k_{z}\right)$, from solution \eqref{GWEFTBTG} and Eq.\eqref{eqdevgeolinear} we obtain the following system of differential equations
\begin{equation}\label{ODESNMPW}
\begin{cases}

\ddot x(t)=\frac{1}{2}\omega_{1}^{2}\left[\hat{\epsilon}^{\left(+\right)}\left(\omega_{1}\right)x+\hat{\epsilon}^{\left(\times\right)}\left(\omega_{1}\right)y\right]e^{i\omega_{1}\left(t-z\right)}+c.c. \\ \\
\ddot y(t)=\frac{1}{2}\omega_{1}^{2}\left[\hat{\epsilon}^{\left(\times\right)}\left(\omega_{1}\right)x-\hat{\epsilon}^{\left(+\right)}\left(\omega_{1}\right)y\right]e^{i\omega_{1}\left(t-z\right)}+c.c.\\ \\
\ddot z(t)=0
\end{cases}\ ,
\end{equation}
whose solutions are the two standard plus and cross polarization modes of GR associated to the frequency $\omega_{1}$, purely transverse, massless of 2-helicity. 

On the other hand, always keeping $\mathbf{k}$ fixed and $k_{m}^{\mu}=\left(\omega_{m},0,0,k_{z}\right)$, with $0\neq k_{m}^{2}=M^{2}_{m}=\omega^{2}_{m}-k^{2}_{z}$ equal to the square of $m$-th mass of scalar field where  $m=2,\dots,n+2$, the $m$-th massive plane wave, from equations  \eqref{eqdevgeolinear} and \eqref{GWEFTBTG}, gives us
\begin{equation}\label{ODESMPW}
\begin{cases}
\ddot x(t)=-\frac{1}{12}\omega^{2}_{m}\hat{B}_{m}\left(k_{z}\right)x e^{i\left(\omega_{m}t-k_{z}z\right)}+c.c.\\ \\
\ddot y(t)=-\frac{1}{12}\omega^{2}_{m}\hat{B}_{m}\left(k_{z}\right)y e^{i\left(\omega_{m}t-k_{z}z\right)}+c.c.\\ \\
\ddot z(t)=-\frac{1}{12}M^{2}_{m}\hat{B}_{m}\left(k_{z}\right)ze^{i\left(\omega_{m}t-k_{z}z\right)}+c.c.
\end{cases}\ .
\end{equation}

Eqs. \eqref{ODESMPW} can be integrated given that we have assumed small the perturbation $h_{\mu\nu}\left(t,z\right)$  and hence we have 
\begin{equation}\label{soluzionseqgeod}
\begin{cases}
x(t)=x(0)+\frac{1}{12}\hat{B}_{m}\left(k_{z}\right)x(0)e^{i\left(\omega_{m}t-k_{z}z\right)}+c.c.\\ \\
y(t)=y(0)+\frac{1}{12}\hat{B}_{m}\left(k_{z}\right)y(0)e^{i\left(\omega_{m}t-k_{z}z\right)}+c.c.\\ \\
z(t)=z(0)+\frac{1}{12\omega^{2}_{m}}M_{m}^{2}\hat{B}_{m}\left(k_{z}\right)z(0)e^{i\left(\omega_{m}t-k_{z}z\right)}+c.c.
\end{cases}\ ,
\end{equation}
where we obtain $n+1$ further mixed massive scalar modes, zero-helicity, partly transverse and partly longitudinal, that is each one with the same mixed scalar polarization. 
In more expressive form,  the linear perturbation of metric $h_{\mu\nu}$,  traveling in the $+\hat{z}$ direction and  keeping  $\mathbf{k}$ fixed may be expressed as 
\begin{equation} 
h_{\mu\nu}\left(t,z\right)=\frac{1}{\sqrt{2}}\left[\hat{\epsilon}^{(+)}\left(k_{z}\right)\epsilon^{(+)}_{\mu\nu}+\hat{\epsilon}^{(\times)}\left(k_{z}\right)\epsilon^{(\times)}_{\mu\nu}\right]e^{i\omega_{1}\left(t-z\right)}+\sum_{m=2}^{n+2}\hat{\epsilon}^{\left(s_{m}\right)}_{\mu\nu}\left(k_{z}\right)e^{i\left(\omega_{m}t-k_{z}z\right)}+c.c.\ ,
\end{equation}
where $\epsilon^{(+)}_{\mu\nu}$ and $\epsilon^{(\times)}_{\mu\nu}$ are the two standard polarization states 

\begin{equation}
\epsilon^{(+)}_{\mu\nu}=\frac{1}{\sqrt{2}}
\begin{pmatrix} 
0 & 0 & 0 & 0 \\
0 & 1 & 0 & 0 \\
0 & 0 & -1 & 0 \\
0 & 0 & 0 & 0
\end{pmatrix}\ ,
\end{equation}
\begin{equation}
\epsilon^{(\times)}_{\mu\nu}=\frac{1}{\sqrt{2}}
\begin{pmatrix} 
0 & 0 & 0 & 0 \\
0 & 0 & 1 & 0 \\
0 & 1 & 0 & 0 \\
0 & 0 & 0 & 0
\end{pmatrix}\ ,
\end{equation}
while $\hat{\epsilon}^{\left(s_{m}\right)}_{\mu\nu}$ is the polarization tensor associated to the scalar modes with $m\in\left\{2,\dots,n+2\right\}$
\begin{equation}
\hat{\epsilon}^{\left(s_{m}\right)}_{\mu\nu}\left(k_{z}\right)=\Biggl[\left(\frac{1}{2}+\frac{\omega^{2}_{m}}{k_{m}^{2}}\right)\epsilon_{\mu\nu}^{(TT)}-\frac{\sqrt{2}\omega_{m}k_{z}}{k_{m}^{2}}\epsilon_{\mu\nu}^{(TS)}\\
-\frac{1}{\sqrt{2}}\epsilon_{\mu\nu}^{(b)}+\left(-\frac{1}{2}+\frac{k_{z}^{2}}{k_{m}^{2}}\right)\epsilon_{\mu\nu}^{(l)}\Biggr]\frac{\hat{B}_{m}\left(k_{z}\right)}{6}\ ,
\end{equation}
with four states
\begin{align}
\epsilon^{(TT)}_{\mu\nu}&=
\begin{pmatrix} 
1 & 0 & 0 & 0 \\
0 & 0 & 0 & 0 \\
0 & 0 & 0 & 0 \\
0 & 0 & 0 & 0
\end{pmatrix}\ , &
\epsilon^{(TS)}_{\mu\nu}&=\frac{1}{\sqrt{2}}
\begin{pmatrix} 
0 & 0 & 0 & 1 \\
0 & 0 & 0 & 0 \\
0 & 0 & 0 & 0 \\
1 & 0 & 0 & 0
\end{pmatrix}\ ,\\
\epsilon^{(b)}_{\mu\nu}&=\frac{1}{\sqrt{2}}
\begin{pmatrix} 
0 & 0 & 0 & 0 \\
0 & 1 & 0 & 0 \\
0 & 0 & 1 & 0 \\
0 & 0 & 0 & 0
\end{pmatrix}\ , &
\epsilon^{(l)}_{\mu\nu}&=
\begin{pmatrix} 
0 & 0 & 0 & 0 \\
0 & 0 & 0 & 0 \\
0 & 0 & 0 & 0 \\
0 & 0 & 0 & 1
\end{pmatrix}\ .
\end{align}
However, only the restriction of the polarization tensor $\hat{\epsilon}^{\left(s\right)}_{\mu\nu}$ to spatial components $\hat{\epsilon}^{\left(s\right)}_{i,j}$ has physical significance 
\begin{equation}
\hat{\epsilon}^{\left(s_{m}\right)}_{i,j}=-\frac{1}{6\sqrt{2}}\hat{B}_{m}\left(k_{z}\right)\epsilon_{i,j}^{(b)}+\frac{1}{6}\left(-\frac{1}{2}+\frac{k_{z}^{2}}{k_{m}^{2}}\right)\hat{B}_{m}\left(k_{z}\right)\epsilon_{i,j}^{(l)}\ ,
\end{equation}
where $(i,j)$ range over $(1,2,3)$. This reduces to two states of polarization, $\epsilon^{(b)}_{\mu\nu}$ and $\epsilon^{(l)}_{\mu\nu}$, that always occur coupled as a combination of the longitudinal scalar mode $\epsilon^{(l)}_{\mu\nu}$ and the transverse breathing scalar mode $\epsilon^{(b)}_{\mu\nu}$.

To visualize gravitational wave polarizations, we use the equation of the geodesic deviations when a $m$-th  gravitational plane wave of frequency $\omega_{m}$ strikes a sphere of freely falling particles of radius $r=\sqrt{x^2(0)+y^{2}(0)+z^{2}(0)}$,  where the displacement of a given particle from the center of the ring $\vec{\chi}=\left(x^1,x^2,x^3\right)$ is given by the solution of the geodesic deviation equation \eqref{soluzionseqgeod}. The sphere will be distorted into an ellipsoid described by
\begin{equation}
\left(\frac{x}{\rho_{1m}(t)}\right)^{2}+\left(\frac{y}{\rho_{1m}(t)}\right)^{2}+\left(\frac{z}{\rho_{2m}(t)}\right)^{2}=r^{2}\ ,
\end{equation}
where both $\rho_{1m}(t)=1+\frac{1}{6}\hat{B}_{m}\left(k_{z}\right)\cos\left(\omega_{m}t-k_{z}z\right)$ and $\rho_{2m}(t)=1+\frac{M_{m}^{2}}{6\omega_{m}^{2}}\hat{B}_{m}\left(k_{z}\right)\cos\left(\omega_{m}t-k_{z}z\right)$ varying between their maximum and minimum value with $m$ range over $\left(2,\dots,n+2\right)$. Each swinging ellipsoid represents an additional scalar mode, zero-helicity which is partly longitudinal and partly transverse \cite{PW}. The d.o.f. of the higher order gravity linear with respect to both $R$ and $\Box^{i} R$ are $n+3$ in all: two of these, $\hat{\epsilon}^{\left(+\right)}$ and $\hat{\epsilon}^{\left(\times\right)}$, generate the $+$ and $\times$ tensor modes and the $n+1$ degree of freedom $\hat{B}_{2}$,..., $\hat{B}_{n+2}$ generate $n+1$ mixed longitudinal-transverse scalar modes . Finally our theory of gravity \eqref{hfr14} shows three polarizations, two transverse tensor polarizations and one mixed transverse-longitudinal scalar polarization, and $n+3$ modes, two tensor modes of frequency $\omega_{1}$ and $n+1$ mixed scalar modes of frequency $\omega_{2},\dots,\omega_{n+2}$. Similarly due to modes spectrum invariance demonstrated earlier also the theory of non-local gravitation \eqref{nlhfr14} exhibits three polarizations and $n+3$ modes with exactly the same properties of the studied case.  However, this is a formal analogy related to the coincidence of the spectrum. The physical meaning of GWs in local and in non-local gravity could be different.

\section{Discussion and Conclusions}
\label{Conclusions}
The number of GW polarizations depends on the considered   theory of gravity. 
In this work,  we have compared oscillation modes of  gravitational waves derived from local  $f \left( R, \Box R, \Box^2 R, \cdots, \Box^n R \right)$ models  with those derived from non-local  $f \left( R, \Box^{-1} R, \Box^{-2} R, \cdots, \Box^{-n} R \right)$ models. Specifically, we have  investigated   the presence of ghost modes and analyzed  the spectrum of scalar modes. The main feature we derived is that  these two classes of  models have exactly the same gravitational  spectrum and therefore  gravitons have formally the same modes and polarizations. This result can be achieved  by noting that, after rewriting the actions by introducing appropriate scalar fields and rescaling the metric, both of them  are invariant under the BRS transformation needed to eliminate ghosts. 

As special cases, we have analyzed  gravitational modes for    $f \left( R, \Box R, \Box^2 R, \cdots, \Box^n R \right) 
= R + \sum_{i=1}^n \alpha_i R \Box^i R$, linear with respect to both $R$ and $\Box^i R$,  and  for  $ f \left( R, \Box R \right) 
= R + \alpha \left(\Box R\right)^2$, quadratic with respect to $\Box R$. 
For $n=1$ or $n=2$, perturbing the action with respect to the flat spacetime, depending on the value of  the constants $\alpha_{i}$, we have obtained canonical scalar fields or ghosts. In the other case, ghost modes depend on the  parameter $\alpha$. 

Finally, we studied GWs in higher order local and non-local gravity comparing the spectrum of the two actions  \eqref{hfr14} and \eqref{nlhfr14}. Using the distribution calculus, we  first solved the linearized field equations associated with  $f \left( R, \Box R, \Box^2 R, \cdots, \Box^n R \right) =R + \sum_{i=1}^n \alpha_i R \Box^i R$ obtaining GWs. They show three polarizations and $n+3$ oscillating modes  related to the  degrees of freedom. The two tensor modes of GR of frequency $\omega_{1}$, both  transversely polarized, massless, with 2-helicity  and  propagation speed $c$ are ruled by the two amplitudes $\epsilon_{\mu\nu}^{\left(+\right)}$ and $\epsilon_{\mu\nu}^{\left(\times\right)}$. In addition to these standard modes, we have obtained $n+1$ scalar modes associated with each frequency $\omega_{m}$ where $m\in(2,\dots,n+2)$. Every scalar mode is ruled by a single degree of freedom $\hat{B}_{m}\left(k_{z}\right)$ and all scalar modes have only one mixed polarization,  partly transversally  and partly longitudinally. Moreover these further  modes are massive with helicity zero and speed of propagation less then $c$.
It is worth noticing that, being the gravitational spectrum of local and non-local gravity formally the same, also GWs modes and polarizations appear  formally the same also if the physical meaning behind them are different. The reason is that the localization procedure of $\Box^{-i}$ operators allows to write actions \eqref{nlhfr13NL} and \eqref{hfr13} in the same way. However, GWs coming from local and non-local gravity have to be physically characterized besides formal aspects.

In this perspective,  the GW170817 event \cite{GW_multi} has been the first reported setting  important constraints and upper bounds on physically viable theories of gravity. In fact,  besides the multi-messenger issues, 
the event  provides constraints on the difference between the speed of electromagnetic  and  gravitational waves.  This fact gives a formidable way to fix the mass of  further gravitational  modes which results very light (see \cite{speed1} for details). 
Furthermore   the GW170817 event allows  the investigation of  equivalence principle (through Shapiro delay measurement) and Lorentz invariance. The limits of possible violations of Lorentz invariance  are reduced by the new observations, by up to ten orders of magnitude \cite{speed1}. This fact is extremely relevant to discriminate between metric and teleparallel formulation of gravitational theories \cite{Bamba, RepT}. Finally,  GW170817 seems to  exclude some alternatives to GR, including classes of  scalar-tensor theories like Brans-Dicke gravity, Horava-Lifshitz gravity, and bimetric gravity  while $f(R)$ gravity seems retained \cite{lombriser}. According to these results, non-local gravity could be not excluded by binary systems observations.

Finally, it seems that  a complete classification of modified theories can be achieved by gravitational waves. However, more events like 
GW170817 are necessary in order to fix precisely gravitational parameters and not giving just upper bounds (see also \cite{Bogdanos,Calmet}). In this context, the present study could constitute a first approach  in order to classify gravitational modes and polarizations pointing out that local and non-local gravity models cannot be simply discriminated by GWs because the "localization" procedure adopted in non-local gravity gives results similar to local gravity.  

Furthermore, it is worth noticing that  primordial gravitational waves in such class of theories have been
investigated \cite{Staro}.  In particular, constraints coming from the CMB on tensor-to-scalar ratio have been considered in   $R^2$-like inflation improved by  non-local modification of gravity.   The power spectrum of tensor perturbations results modified due to the non-local Weyl tensor squared term. This is a strong indication that  future CMB data can probe non-local behavior of gravity at high space-time curvatures. In other words, a combined multimessenger and cosmological approach could definitely probe non-local gravity.

In a forthcoming paper,  GWs solutions, in the framework of non-local gravity, will be  derived adopting the distribution calculus.

%%%%%%%%%%%%%%%%%%%%%%%%%%%%%%%%%%%%%%%%%%%%%%%
\section*{Acknowledgements}
%%%%%%%%%%%%%%%%%%%%%%%%%%%%%%%%%%%%%%%%%%%%
SC wants to thank the  Kobayashi-Maskawa Institute for the Origin of Particles and the Universe for the kind hospitality during August 2018 when these discussions started.
SC is supported in part by the INFN sezione di Napoli, {\it iniziative specifiche} MOONLIGHT2 and QGSKY. 

%%%%%%%%%%%%%%%%%%%%%%%%%%%%%%%%%%%%%%%%%%%%%%%%%

\end{document}